\documentclass[runningheads]{llncs}

\usepackage[colorlinks,
linkcolor=blue,
anchorcolor=blue,
citecolor=blue]{hyperref}

\usepackage{graphicx}
\usepackage{amsmath,amssymb,amsfonts}
\usepackage{textcomp}
\usepackage{makecell}
\usepackage{booktabs}
\usepackage{threeparttable}

\begin{document}
%
%\title{Cross-architecture Binary Similarity Comparison with Multi-relational Instruction Association Graph \\}
\title{Multi-relational Instruction Association Graph for Cross-architecture Binary Similarity Comparison}
%
%\titlerunning{Cross-architecture Binary Similarity Comparison} 
\titlerunning{Multi-relational Instruction Association Graph for Binary Comparison} 
% If the paper title is too long for the running head, you can set
% an abbreviated paper title here
%
\author{Qige Song\inst{1, 2} \and
	Yongzheng Zhang\inst{3} \and
	Shuhao Li\inst{2}}
\authorrunning{Qige Song et al.}
% First names are abbreviated in the running head.
% If there are more than two authors, 'et al.' is used.
%
\institute{School of Cyber Security, University of Chinese Academy of Sciences, Beijing, China \and
	Institute of Information Engineering, Chinese Academy of Sciences, Beijing, China
	\email{songqige@iie.ac.cn, lishuhao@iie.ac.cn} \and
	China Assets Cybersecurity Technology CO.,Ltd., Beijing, China\\
	\email{zhangyz@cacts.cn}}
\maketitle              % typeset the header of the contribution

\begin{abstract}
Cross-architecture binary similarity comparison is essential in many security applications. Recently, researchers have proposed learning-based approaches to improve comparison performance. They adopted a paradigm of instruction pre-training, individual binary encoding, and distance-based similarity comparison. However, instruction embeddings pre-trained on external code corpus are not universal in diverse real-world applications. And separately encoding cross-architecture binaries will accumulate the semantic gap of instruction sets, limiting the comparison accuracy. This paper proposes a novel cross-architecture binary similarity comparison approach with multi-relational instruction association graph. We associate mono-architecture instruction tokens with context relevance and cross-architecture tokens with potential semantic correlations from different perspectives. Then we exploit the relational graph convolutional network (R-GCN) to perform type-specific graph information propagation. Our approach can bridge the gap in the cross-architecture instruction representation spaces while avoiding the external pre-training workload. We conduct extensive experiments on basic block-level and function-level datasets to prove the superiority of our approach. Furthermore, evaluations on a large-scale real-world IoT malware reuse function collection show that our approach is valuable for identifying malware propagated on IoT devices of various architectures.

\end{abstract}

\section{Introduction}
Cross-architecture binary similarity comparison task aims at measuring the functional semantic similarity of binary snippets compiled from different CPU architectures. It is of great significance in many systems security applications, such as vulnerability detection, patch analysis, and malware detection. In this paper, we explore its application in malware defense of the Internet of Things (IoT) environments. IoT devices have been widely used in various real-life scenarios in recent years. However, the security protection of many IoT devices is not yet perfect, leaving hidden dangers such as weak authentication services and security vulnerabilities, which have attracted the attention of malware developers. Many attackers use malware as a weapon to invade vulnerable devices to build large-scale IoT botnets and operate them to launch Distributed Denial of Service (DDoS) attacks, causing severe damage \cite{antonakakis2017understanding,costin2018iot,herwig2019measurement}. 

Due to the diversity of underlying hardware architectures of IoT devices, malware developers often reuse source code to generate malware binaries of multiple architectures, thereby infecting more devices and expanding the scale of attacks. Cozzi \emph{et al.} \cite{cozzi2020tangled} conducted empirical analysis on over 93k IoT malware emerged between 2015 and 2018, and results have shown that the samples involve more than a dozen kinds of architectures, with \emph{ARM} and \emph{MIPS} in the majority. They also found the prevalence of code reuse across malware samples of different families and architectures. Wang \emph{et al.} \cite{wang2020iotcmal} deployed IoT honeypots in real-world network environments and captured a large number of wild IoT malware samples. After code-level analysis, they confirmed that many samples with the same code origin are reused and propagated on IoT devices of different architectures. Therefore, an effective cross-architecture binary code similarity comparison approach can help discover reused IoT malware fragments under different architectures, providing solutions for defending against IoT attacks.

Traditional binary similarity comparison approaches are categorized into two main technical routes, static analysis-based approaches and dynamic analysis-based approaches. The former compares the syntactic or statistical features of disassembly instruction sequences \cite{hu2013mutantx,khoo2013rendezvous,ng2013expose}, or design hash algorithms to calculate the similarity of binary fragments \cite{farhadi2014binclone,qiao2016fast,xu2017spain}. However, since different architectures have separate instruction sets, the mnemonics, registers, and memory access strategies are different, making it difficult to achieve decent performance. Dynamic-based approaches compare the runtime state information \cite{kargen2017towards} or the input-output pairs of binary code fragments to measure their semantic similarity \cite{david2016statistical,ming2017binsim}. But it is challenging to perform scalable analysis on large-scale binary collections. Meanwhile, supporting diverse architectures and compilation settings will lead to extra workload for dynamic analysis environment configuration.

Considering the above disadvantages, researchers have recently shifted their focus to learning-based approaches. They generally adopted the following working paradigm: (\romannumeral1) Generate the initial representations of disassembly instructions by manually-designed features or external pre-training mechanism. (\romannumeral2) Exploit deep neural encoders to individually extract vectorized representations of each binary snippet. (\romannumeral3) Calculate the similarity score based on the distance metrics. Specifically, \texttt{Gemini} \cite{xu2017neural} extracted statistical vectors of basic blocks within the control-flow-graph of binary function pairs, then encoded the overall graph with the \emph{structure2vec} network and measured their cosine similarity. \texttt{INNEREYE} \cite{zuo2018neural} used \emph{word2vec} to learn the assembly instructions embeddings (numerical vectors) on external code corpus, then deployed RNN encoders to separately generate sequential features of instruction sequences, and evaluated the similarity of cross-architecture binary snippets based on the \emph{Manhattan} distance. \texttt{SAFE} \cite{massarelli2019safe} set up similar instruction representation and sequence encoding modules, with additional self-attention layers to automatically assign high weights to important instructions, improving the semantic similarity matching performance. Although showing promising results, these approaches have two main dilemmas:

\begin{itemize}
	\item First, these approaches individually encode binaries of different architectures and then perform similarity comparisons on their vectorized representations. However, the semantic representation spaces of the cross-architecture instruction sets have a clear gap. The separate binary characteristics encoding mechanism will accumulate this gap, resulting in inaccurate comparison results. This phenomenon will be more pronounced for larger binary snippets.
	
	\item Second, although state-of-the-arts have proved that pre-trained instruction representations perform better than statistical features \cite{zuo2018neural,redmond2018cross,yu2020order}, it relies on high-quality, large-scale external code corpora. Considering real-world binaries are generated under diverse architectures and compilation settings, especially for IoT environments, collecting a comprehensive corpus for instruction pre-training generally takes considerable time and effort.
\end{itemize}

%To tackle the aforementioned challenges,
In this paper, we design a novel cross-architecture binary similarity comparison approach with \emph{multi-relational instruction association graph}, which can effectively alleviate these two deficiencies. For a pair of binary snippets, we first associate mono-architecture opcode and operand token pairs with the operational relationship or co-occurrence-based relevance. Then we associate the cross-architecture token pairs with multi-perspective potential semantic correlations, including prefix-match, value-match, type-match and heuristic position alignment dependencies. The mono-architecture associations can effectively enhance the functional semantic representation of instructions without external instruction pre-training. The cross-architecture associations can effectively bridge the gap in the semantic feature spaces of instructions under different architectures, and significantly improve the cross-architecture binary similarity matching performance. We apply the relational graph convolutional network (R-GCN) to propagate information on the constructed multi-relational instruction association graph. R-GCN groups different types of relations and uses a separate neighbor message-passing mechanism, which can effectively aggregate multi-type associations and iteratively refine the semantic representations of instruction tokens.

%We implement our solution as an end-to-end cross-architecture binary similarity comparison framework. We first generate the initial instruction representation and sequence representation of the input binary snippets pair based on an instruction vectorization module and an instruction sequence encoding module. Then we construct the corresponding multi-relational instruction association graph for each pair and use stacked R-GCN layers to obtain the improved instruction representation and sequence representation. Finally, we fuse the refined representation vector of the binary snippets pair and generate their similarity comparison score.

Our major contributions can be summarized as follows:
\begin{itemize}
	\item We creatively design a novel cross-architecture similarity binary comparison approach with multi-relational instruction association graph. It can associate the semantic dependencies of instructions from different perspectives, bridge the gap in the representation spaces of cross-architecture instructions, and significantly improve binary comparison performance.
	
	\item We implement our solution as an end-to-end cross-architecture binary similarity comparison system and conduct extensive evaluations at two granularities, basic block-level and function-level. Results show that our approach significantly outperforms the existing learning-based approaches (AUC=0.9924 for basic block-level and Precision@1 = 0.9216 for function-level).
	%extensive experimental study
	%Results show that our approach significantly outperforms the state-of-the-arts (1.64\% improvement for basic block-level dataset and 29.86\% improvement for function-level dataset).
	
	\item We further evaluate our approach on a large-scale cross-architecture reuse function dataset (460,386 pairs) constructed from IoT malware in real-world environments. Promising results prove that our method is valuable for defending against malware spread on IoT devices of different architectures.
\end{itemize}

\section{Overview}
\subsection{Problem Statement}
%\noindent \textbf{Problem Statement.}
In this section, we give a formal definition of the cross-architecture binary code similarity comparison problem. The input is a pair of binary code snippets compiled on two different CPU architectures, and the output is a semantic similarity score. We set the output value in the range of 0 to 1, where 1 represents that the input binary snippets are compiled from the exact same source code, and 0 means that their source code implements completely different functions.

\subsection{System Workflow}
%\noindent \textbf{System Workflow.}
Our approach is implemented as an end-to-end cross-architecture binary code similarity comparison system. Figure \ref {workflow} shows the overall system workflow, including four major modules:

\begin{figure*}[!t]
	\centering
	\includegraphics[width=10cm]{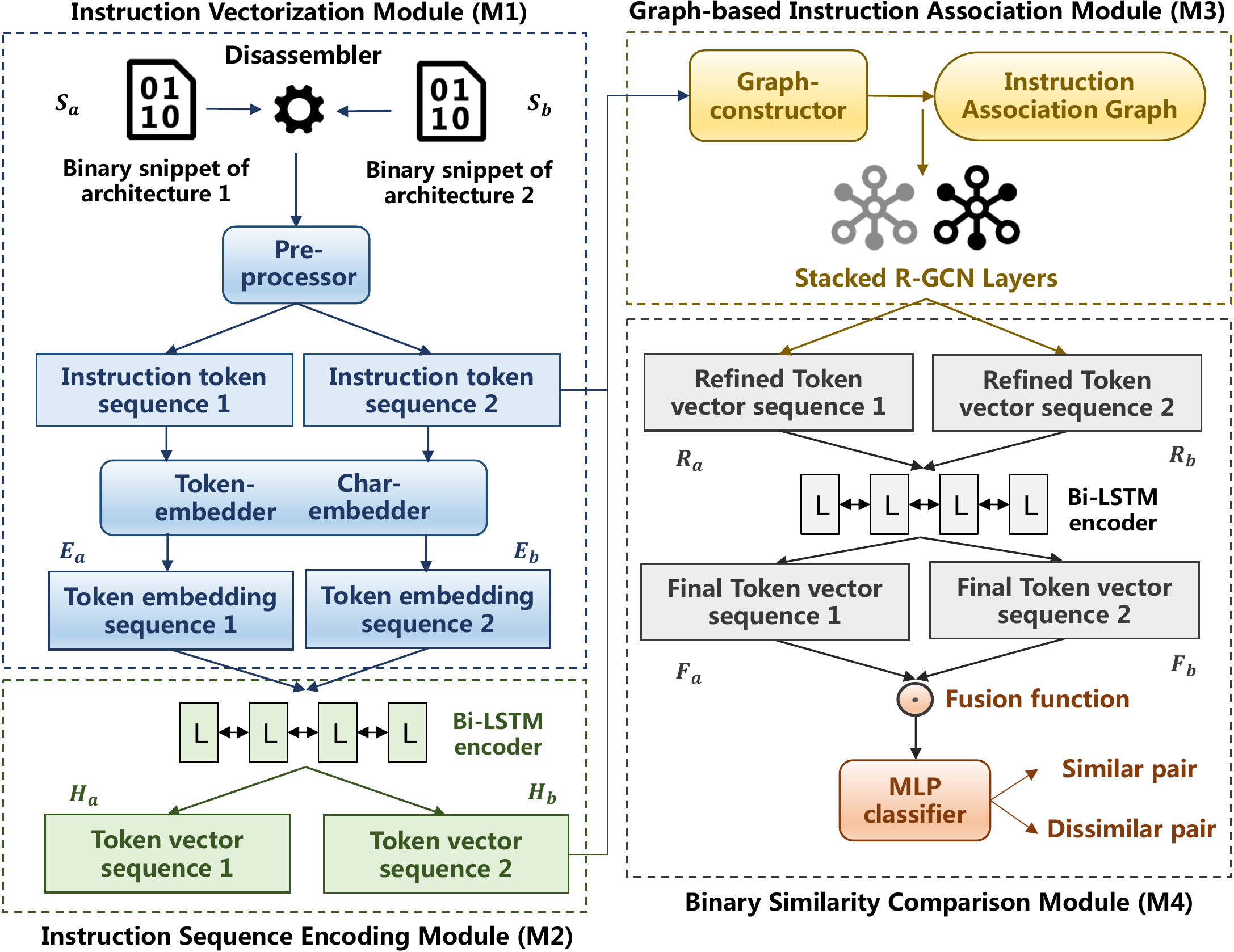}
	\caption{The workflow of our proposed multi-relational instruction association graph-based cross-architecture binary similarity comparison approach. M1 is the abbreviation of \emph{Module 1}.}
	\label{workflow}
	%\vspace*{-0.8\baselineskip}
\end{figure*}

\begin{itemize}
	\item Instruction representation module (\emph{Module 1}): We disassemble and preprocess the input binary snippets, obtaining the initial instruction token representations with token-level and character-level (char-level) features.
	
	\item Instruction sequence encoding module (\emph{Module 2}): We use Bi-LSTM as the backbone encoder to extract the sequential representations of disassembly instruction token sequences.
	
	\item Graph-based instruction association module  (\emph{Module 3}): We construct the corresponding instruction association graph of the input pair of binary snippets, then generate the refined token representations by relational graph convolutional network.
	
	\item Binary similarity comparison module  (\emph{Module 4}): We fuse the final representation vectors of the binary snippets and generate the similarity comparison score by Multilayer Perceptron (MLP) network.
\end{itemize}

\section{Instruction Vectorization Module}
The input pair of binary snippets $S_a$ and $S_b$ are originally discrete byte streams, and we first extract their corresponding disassembly instruction sequences. Each instruction contains an opcode and a set of operands. The opcode specifies the operation performed by the instruction, and the operands represents the operation object, like registers and immediate literals. We tokenize raw disassembly instruction sequences and treat each independent opcode or operand as a token unit for vectorization. The input pair can be represented as token sequence $T_a$ = ($t_{a_1}$, $t_{a_2}$, ..., $t_{l_a}$), and $T_b$ = ($t_{b_1}$, $t_{b_2}$, ..., $t_{l_b}$). $l_a$ and $l_b$ are the length of the sequences. To improve generality, we preprocess the original token sequence, replacing the numerical constants with 0 while preserving the negative signs.

\begin{figure}[!t]
	\centering
	\includegraphics[width=12cm]{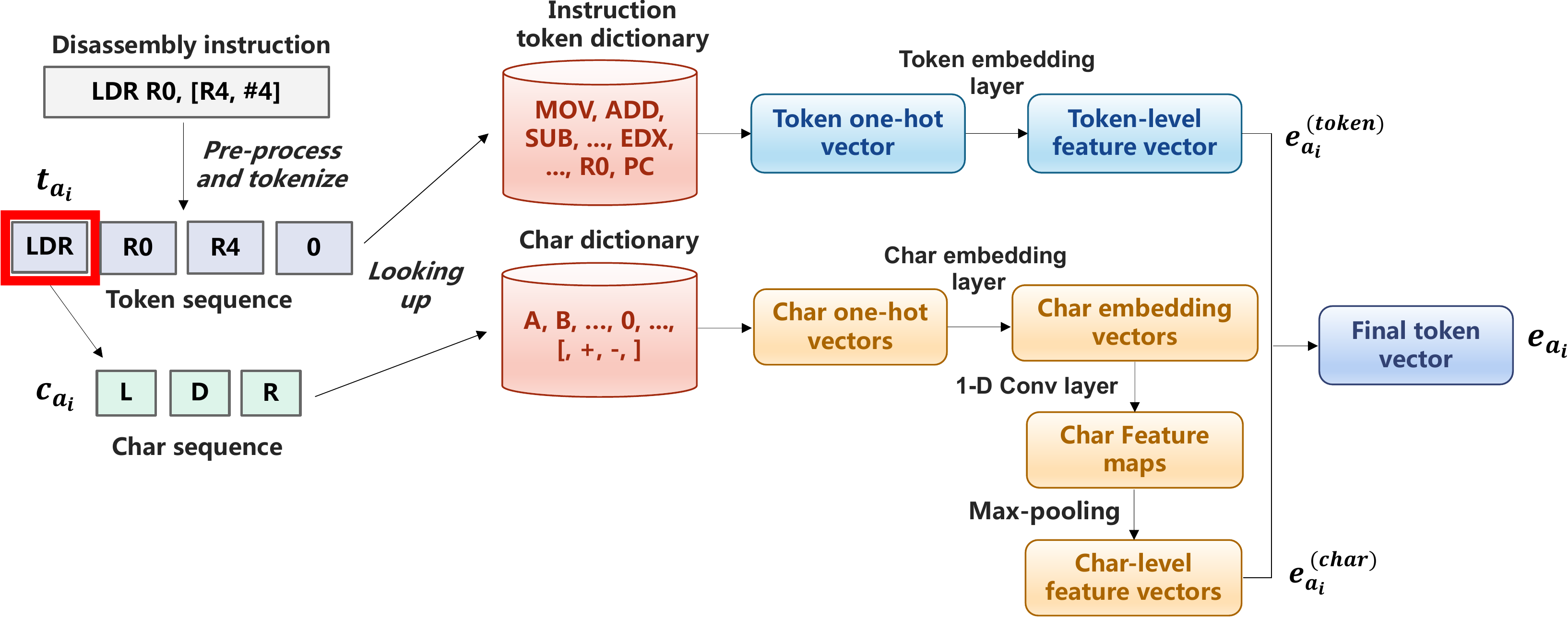}
	\caption{Detailed instruction vectorization process.}
	\label{instruction_representation}
	%\vspace*{-0.8\baselineskip}
\end{figure}

Figure \ref {instruction_representation} shows the working process of our instruction token vectorization module. We create a token lookup dictionary for the processed opcode and operand units. The initial representation of each token is a sparse one-hot vector based on its index within the dictionary, and then we set a trainable token embedding layer to learn the discrete dense vector $\mathbf{e}^{(token)}_{a_i}$ of token $t_{a_i}$.

We further extract the char-level features to enrich the instruction token representations with lexical information. We extract the corresponding character (char) sequences of the tokens and generate a char lookup dictionary. Then each token can be represented as a sequence of one-hot vectors of the chars. Similar to the token embedding process, we set up an embedding layer to obtain the dense vector of each char. After that, we deploy a one-dimensional-Convolutional (1D-Conv) layer to extract the local-spatial features of the char sequences. It will slide fixed-size 1D-Conv filters over the char vector sequence of the instruction token and generate corresponding feature maps. Then we process the feature maps by a max-pooling layer to generate the final char-level representations. In specific, for token $t_{a_i}$ with char one-hot sequence ($\mathbf{c}_{{a_i}_1}$, $\mathbf{c}_{{a_i}_2}$, ..., $\mathbf{c}_{{a_i}_M}$) of length $M$, its char-level vectorized representation is generated as follows:

\begin{equation}
	\mathbf{e}^{(char)}_{a_i}=\operatorname{max-pooling} \left(\operatorname{1D-Conv}\left(\mathbf{c}_{{a_i}_1}, \mathbf{c}_{{a_i}_2}, ..., \mathbf{c}_{{a_i}_M}\right)\right)
\end{equation}

The final vectorization result $\mathbf{e}_{{a_i}}$ of the instruction token $t_{a_i}$ is the concatenation of token embedding features and character-level features. Note that our initial instruction vectorization process is implemented entirely through learnable parameters without relying on any external pre-trained instruction embeddings.

\begin{equation}
	\mathbf{e}_{{a_i}}=\operatorname{Concat}\left(\mathbf{e}^{(token)}_{a_i} ; \mathbf{e}^ {(char)}_{a_i}\right)
\end{equation}

\section{Instruction Sequence Encoding Module}
From the instruction vectorization module, we obtain the instruction token vector sequences $\mathbf{E}_{a}$ = ($\mathbf{e}_{{a_1}}$, $\mathbf{e}_{{a_2}}$, ..., $\mathbf{e}_{{l_a}}$) and $\mathbf{E}_{b}$ = ($\mathbf{e}_{{b_1}}$, $\mathbf{e}_{{b_2}}$, ..., $\mathbf{e}_{{l_b}}$) of binary snippets $S_a$ and $S_b$. Our next goal is to encode the functional semantics of the overall binary sequences and generate their meaningful representations. The recurrent neural network (RNN) has a strong sequential context modeling ability and has been widely used in text sequence characterization. We apply it into our binary instruction sequence encoding process. To prevent the gradient vanishing and exploding problems that are prone to occur when encoding long sequences, we apply the LSTM variant instead of the vanilla RNN. We deploy the bidirectional LSTM (Bi-LSTM), with two LSTMs separately encoding forward and backward information. The hidden state $\mathbf{h}_{a_i}$ of token $t_{a_i}$ is generated as follows:

\begin{equation}
	\overrightarrow{\mathbf{h}_{a_i}}=\overrightarrow{\operatorname{LSTM}}\left(t_{a_1}, t_{a_2}, \ldots, t_{a_i}\right), \overleftarrow{\mathbf{h}_{a_i}}=\overleftarrow{\operatorname{LSTM}}\left(t_{l_a}, t_{l_a - 1}, \ldots, t_{a_i}\right)
\end{equation}

\begin{equation}
	\mathbf{h}_{a_i}=\operatorname{Concat}\left(\overrightarrow{\mathbf{h}_{a_i}}; \overleftarrow{\mathbf{h}_{a_i}}\right)
\end{equation}

After the encoding layers, the binary snippets are represented as the sequences of token hidden vectors, $\mathbf{H}_a$ =  ($\mathbf{h}_{{a_1}}$, $\mathbf{h}_{{a_2}}$, ..., $\mathbf{h}_{{l_a}}$), $\mathbf{H}_b$ = ($\mathbf{h}_{{b_1}}$, $\mathbf{h}_{{b_2}}$, ..., $\mathbf{h}_{{l_b}}$). We keep all tokens' hidden states without any aggregation or pooling operation. They will be used for subsequent instruction association and instruction representation refinement process.

\section{Graph-based Instruction Association Module}
\subsection{Multi-relational Instruction Association Graph}
In this section, we give a specific description of our designed instruction association graph schema. Our goal is to model dependencies of instructions from multiple perspectives and refine the quality of instruction token representations. For the instruction token sequence $T_a$ and $T_b$ processed from the input binary snippets, we regard each token as a node and establish the corresponding instruction association graph $\mathcal{G}=(\mathcal{T}, \mathcal{E}, \mathcal{R})$. $\mathcal{T}$ is the node set, $\mathcal{E}$ denotes the edge set, and $\mathcal{R}$ is the type set of the edges. Note that we construct an independent instruction association graph for each binary pair to be compared, so the graph structure will change dynamically for different input pairs.

We design the following six types of edges to represent the semantic relationships between instruction token nodes, including edges of mono-architecture token pairs and edges of cross-architecture token pairs.

\subsubsection{Mono-architecture Association Edges}
\begin{itemize}
	\item  \textbf{Mono-architecture opcode-operate-operand edge ($e_0$)} : We associate the opcode with each operand within the same disassembly instruction, indicating the operational relationship.
	
	\item \textbf{Mono-architecture operands co-occurrence edge ($e_1$)} : We associate each operand pair within a disassembly instruction, displaying their co-occurrence-based relevance.
\end{itemize}

In specific, for the instruction ``$MOV\sim R0, R4$'', we associate token ``$MOV$'' with tokens ``$R0$'' and ``$R4$'' by edges of type $e_0$, indicating that opcode ``$MOV$'' operates on registers ``$R0$'' and ``$R4$''. Meanwhile, tokens ``$R0$'' and ``$R4$'' are associated with the $e_1$ edge, denoting that they co-occurred within the same disassembly instruction. These two types of edges will establish context-based dependencies of the mono-architecture token sequence, improving the semantic representations of opcodes and operands without resorting to external instruction pre-training.

\subsubsection{Cross-architecture Association Edges}
\begin{itemize}
	\item \textbf{Cross-architecture opcodes prefix-match edge ($e_2$)} : We connect two cross-architecture opcodes with the same $n$ prefix characters. Although different architectures have separate instruction sets, some opcodes that perform similar operations have similar char-level lexical characteristics. Such as ``\emph{SUB}'', ``\emph{SUBSD}'', ``\emph{SUBPD}'', ``\emph{SUBSS}'' of the \emph{x86} architectures, and ``\emph{SUB}'', ``\emph{SUBS}'' of the \emph{ARM} architecture will perform similar operations, and they all contain the ``\emph{SUB}'' prefix. This type of edge is meaningful for identifying instruction sequences implementing similar functions.
	
	\item\textbf{Cross-architecture operands value-match edge ($e_3$)} : We associate two cross-architecture operands with the same value after preprocessed. It will establish dependencies between numeric constants, identical symbols, and identical string literals of different architectures, assisting the semantic comparison process of disassembled code fragements.
	
	\item \textbf{Cross-architecture operands type-match edge ($e_4$)} : For a similar purpose to $e_3$, we associate cross-architecture operands with the same fine-grained category. We consider three fine-grained operand types: registers, immediate literals, and memory address pointers.
	%numeric constant values.
	
	\item \textbf{Cross-architecture heuristic position alignment edge ($e_5$)} : We define heuristic rules to establish positional associations between the token pairs of cross-architecture instruction sequences. Specifically, tokens $t_{a_i}$ and $t_{b_j}$ of sequences $T_a$ and $T_b$ will be associated if they meet the following conditions:
	
	\begin{equation}
	|\frac{a_i \times l_a}{l_b} - b_j| < \iota
	\end{equation}
	$a_i$ and $b_j$ are the position indices of the tokens in the respective sequences. $l_a$ and $l_b$ are the lengths of the  sequences. $\iota$ is a pre-defined threshold.
\end{itemize}

The $e_5$ edge is inspired by Redmon \emph{et al.} \cite{redmond2018cross}. We set the edge weights of the instruction association graph as their statistical frequencies, $e_0$ is unidirectional, and $e_1$ to $e_5$ are undirected edges. The same endpoint token nodes may be associated with multiple different types of edges, and our strategy of utilizing multi-type of dependencies will be illustrated in the next subsection.

\subsection{Relational Graph Convolutional Network}
After building the instruction association graph, we expect to refine the semantic representation of instruction tokens with the multi-relational graph schema. Graph neural network (GNN) \cite{scarselli2008graph} has strong graph structures representation abilities, and it has achieved promising performance in diverse tasks such as node classification and link prediction. The classic GNN is a message-passing framework with two core operations, neighbor node information aggregation and node representation update. It iteratively aggregates neighbor node information and updates the vectorized representations of nodes with local subgraph features.

Specifically, for a token node $t_v$ of set $\mathcal{T}$ with the corresponding neighbor set $\mathcal{N}(t_v)$, the neighbor node information aggregation and node representation update process of a GNN layer are performed as:

\begin{equation}
	\mathbf{h}_{t_v}=\mathcal{F}\left(\mathbf{x}_{t_v}, AGG\left(\left\{\mathbf{h}_{t_u}: t_u \in \mathcal{N}(t_v)\right\}\right)\right)
\end{equation}
$\mathbf{x}_{t_v}$ is the initial representation of $t_v$, and $\mathbf{h}_{t_v}$ denotes the updated representation calculated by the aggregated neighbor node features and the fusion function $\mathcal{F}$.

Our instruction association graph connects instruction token pairs with semantically distinct edges. To better exploit the multi-type dependencies, we choose the relational graph convolutional network (R-GCN) \cite{schlichtkrull2018modeling}, a GNN variant that can efficiently handle multi-relational graph data. R-GCN extends the classic GNN framework with type-specific parameters to model the relations of different types separately. For token node $t_v$, the $l^{th}$ R-GCN layer maps its neighbor node-set $ \mathcal{N}_{(t_v)}^{e_r}$ associated with edge type $e_r$ to a unified vector space through relation-specific transformation matrix $W_{e_r}^{l}$, and aggregates the transformed vectors in the normalized summation way. The node's characteristics of the previous layer $\mathbf{h}_{t_v}^{l}$ is preserved by a separate feature transformation matrix $W_{0}^{l}$. The overall $l^{th}$ layer node representation refinement process is as follows:

\begin{equation}
	\mathbf{h}_{t_v}^{l+1}=\sigma\left(\sum_{e_r \in \mathcal{R}} \sum_{t_u \in \mathcal{N}_{t_v}^{e_r}} \frac{1}{|\mathcal{N}_{t_v}^{e_r}|} W_{e_r}^{l} \mathbf{h}_{t_u}^{l}+W_{0}^{l} \mathbf{h}_{t_v}^{l}\right)
\end{equation}

A Single R-GCN layer can only aggregate first-order neighbor information. We stack multiple R-GCN layers to propagate multi-hop neighbor messages, learning the substructure of different scales within the instruction association graph. 
For token $t_{a_i}$, its corresponding input node attributes of the first R-GCN layer are the hidden states $\mathbf{h}_{a_i}$ generated by the previous token vectorization layers and the Bi-LSTM encoder, which contains the tokens' lexical semantics and the sequential information of the mono-architecture binary sequence. The final R-GCN layer will output its refined representation $\mathbf{r}_{{a_i}}$, and $B_a$ and $B_b$ will be represented as $\mathbf{R}_{a}$ = ($\mathbf{r}_{{a_1}}$, $\mathbf{r}_{{a_2}}$, ..., $\mathbf{r}_{{l_a}}$) and $\mathbf{R}_{b}$ = ($\mathbf{r}_{{b_1}}$, $\mathbf{r}_{{b_2}}$, ..., $\mathbf{r}_{{l_b}}$).

\begin{figure*}[!t]
	\centering
	\includegraphics[width=11cm]{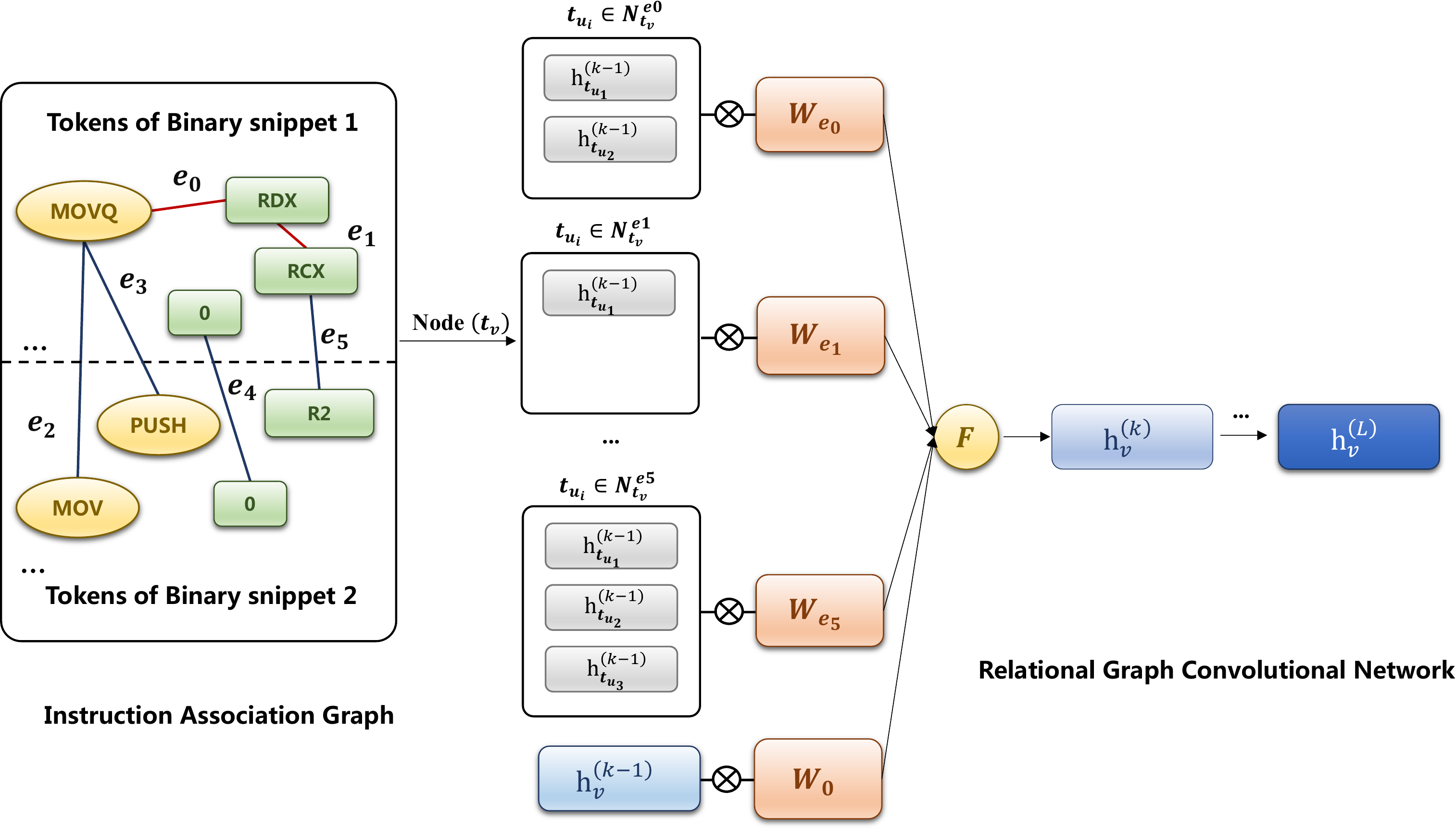}
	\caption{Refine instruction representation with instruction association graph and R-GCN layers.}
	\label{association_graph}
	%\vspace*{-0.8\baselineskip}
\end{figure*}

Figure \ref{association_graph} is the technical explanation of the overall graph-based instruction association module. The left side is an instruction association graph schema corresponding to an input pair of disassembly instruction sequences. We reserve one example edge for each type. On the right is the node representation refinement process performed by R-GCN layers.

\section{Binary Similarity Comparison Module}
The graph-based instruction association module will generate refined instruction token representation sequences $\mathbf{R}_{a}$ and $\mathbf{R}_{b}$. We send the sequences to another Bi-LSTM layer to strengthen the sequential context information, and then employ a max-pooling layer to generate the final binary snippet representations $\mathbf{F}_a$, $\mathbf{F}_b$. The process can be simplified as follows:

\begin{equation}
	\mathbf{F}_{a}=\operatorname{max-pooling}\left(\operatorname{Bi-LSTM}\left(\mathbf{r}_{{a_1}}, \mathbf{r}_{{a_2}}, ..., \mathbf{r}_{l_a}\right)\right)
\end{equation}

%\begin{equation}
%	\mathbf{F}_{b}=\operatorname{max-pooling}\left(\operatorname{Bi-LSTM}\left(\mathbf{r}_{{b_1}}, \mathbf{r}_{{b_2}}, ..., \mathbf{r}_{l_b}\right)\right)
%\end{equation}

We set up a fusion function to aggregate the final vector $\mathbf{F}_a$ and $\mathbf{F}_b$, and then feed the result into a one-layer MLP classifier with the $softmax$ function to predict the similarity score of $S_a$ and $S_b$. The fusion function is based on \cite{conneau2017supervised}, including the concatenation, element-wise difference, and element-wise product of $\mathbf{F}_a$ and $\mathbf{F}_b$, as follows:

\begin{equation}
	\mathbf{F}_{(a, b)}=\operatorname{Concat}\left(\mathbf{F}_{a} ; \mathbf{F}_{b} ; \mathbf{F}_{a}-\mathbf{F}_{b} ; \mathbf{F}_{a} \odot \mathbf{F}_{b}\right)
\end{equation}
This fusion way ensures that the MLP classifier can identify the boundary of the two representation vectors, and calculate the similarity between $F_a$ and $F_b$ more accurately. The overall neural model is optimized by minimizing the cross-entropy loss between the predicted scores and the ground truth pair labels.

\section{Evaluation}
\subsection{Preliminary}
\subsubsection{Dataset.}
We evaluate our approach on three datasets with two granularities:

(\romannumeral1) \textbf{Dataset1} is a collection of basic block pairs provided by \texttt{INNEREYE} \cite{innereye_data} with annotated ground truth, compiled from \texttt{coreutils}-8.29, \texttt{findutils}-4.6.0, \texttt{diffutils}-3.6, \texttt{binutils}-2.30, and \texttt{OpenSSL}-1.1.1-pre1 packages by the \texttt{clang}-6.0.0 compiler and O2 optimization, into \emph{x86} and \emph{ARM} architectures.

(\romannumeral2) We build the function-level \textbf{Dataset2} from the same packages of Dataset1 by two compilers, \texttt{clang}-6.0.0 and \texttt{GCC}-5.4.0, also with O2 optimization.

(\romannumeral3) We construct \textbf{Dataset3} with seven families of IoT malware samples captured by honeypots deployed in real-world network environments. We select samples spread on \emph{ARM} and \emph{MIPS} architectures, which are widely used in IoT devices, and build a large-scale cross-architecture malware reuse function matching dataset. 

\begin{table*}
	\caption{Statistical information of evaluation datasets}
	%\begin{threeparttable}
	\label{dataset_info}
	\centering
	\begin{tabular}{m{50pt}<{\centering}m{58pt}<{\centering}m{55pt}<{\centering}m{55pt}<{\centering}m{60pt}<{\centering}m{60pt}<{\centering}}
		%\begin{tabular}{ccccc}
		\toprule
		Dataset & Architectures & Granularity & \# Pairs & Average \# Edges & Average \# Instructions \\
		
		\midrule
		Dataset1 & \emph{x86-ARM} & Basic-Block & 11,2019 & 239.50 & 7.09 \\
		Dataset2 &  \emph{x86-ARM } & Function & 74,841 & 1142.31 & 28.38 \\
		%Dataset3 &  \emph{ARM-MIPS} & Function & 460,386 & 876 \\
		%1,563
		
		Dataset3 &  \emph{ARM-MIPS} & Function & 460,386 & 737.21 & 33.49 \\
		
		\bottomrule
	\end{tabular}
\end{table*}

We annotate the ground truth of Dataset2 and Dataset3 with binary name and function name as the unique ID to identify functions compiled from the same source code. When building Dataset3, we filtered out functions of stripped binaries whose names could not be correctly identified by the disassembler. Table \ref{dataset_info} shows the information of our datasets. The last two columns represent the average number of edges of the instruction-association graphs, and the average number of disassembly instructions contained in each binary snippet.

\subsubsection{Evaluation metrics.}
We consider two groups of metrics in evaluation:
 
(\romannumeral1)  For basic block-level evaluation, we follow the prior art \cite{zuo2018neural,redmond2018cross} and use AUC-ROC (Area under ROC curve) as the evaluation metric.

(\romannumeral2) For function-level evaluation, we set up the function search task as the prior art \cite{liu2018alphadiff,massarelli2019safe,yu2020order}. It will compare a query binary function with multiple candidate functions and rank the corresponding similarity scores. In our experiments, we set one positive candidate similar to the query function, together with $N_{neg}$ dissimilar candidates. We adopt two commonly used evaluation metrics, precision@1 and MRR (Mean Reciprocal Rank). For each query function, the precision@1 value indicates whether the score of the positive candidate function is ranked first, and the MRR value is calculated as the reciprocal of the positive candidate's ranking position.

\subsubsection{Implementation details.}
We implement the proposed method with the \emph{PyTorch} framework, \emph{PyTorch Geometric} library \cite{FeyLenssen2019}, and \emph{radare2} disassembler \cite{radare2}. The parameters are optimized by Adam with a learning rate of 1e-3. The dimension of the instruction token embedding layer is set to 128, and the char embedding dimension is 32. The size of char 1-D convolutional filters is 2, and the number is 64. We set up one Bi-LSTM layer and two R-GCN layers with the hidden dimension of 256. The $n$ value of the \emph{Cross-architecture opcode prefix-match edge} ($e_2$) is set to 3, and the $\iota$ threshold of the  \emph{Cross-architecture heuristic position alignment edge} ($e_5$) in the equation 5 is set to 2. Under this setting, our system can achieve the best results on the development sets. For Dataset2 and Dataset3, the number of negative candidates $N_{neg}$ for each query function is 20.

\subsection{Basic block-level Experiments}
\subsubsection{Comparisons with the prior art.}
In this section, we evaluate our approach at the basic block-level dataset and compare it with two manually designed baselines and three state-of-the-art cross-architecture binary similarity comparison approaches. The implementations of the comparison methods are as follows:

\begin{itemize}
	\item String edit distance: We use \texttt{python}-\texttt{Levenshtein} \cite{edit_distance}  to compute edit-distance-based similarity scores of the cross-architecture basic block pairs.
	
	\item Char n-gram: We extract the char n-gram sets of the assembly instruction sequences and calculate their Jaccard similarity score. 4-gram performs best in our evaluation.
	
	\item \texttt{Gemini} \cite{xu2017neural} features + SVM: Xu \emph{et al.} extracted the statistical features to represent a basic block, such as the number of instructions, calls, and numeric constant. Following Zuo \emph{et al.} \cite{zuo2018neural}, we concatenate the features of two blocks and use SVM with \texttt{RBF} kernel to predict their similarity.
	
	\item \texttt{INNEREYE} \cite{zuo2018neural}: Zuo \emph{et al.} used \emph{word2vec} to pre-train assembly instruction embeddings on external code corpus. Then encode the cross-architecture instruction sequences by two individual LSTM networks and generate the similarity score based on the distance metric.
	
 	%LSTM \cite{zuo2018neural}: \texttt{INNEREYE} used \emph{word2vec} to pre-train assembly instruction embeddings on external code corpus. Then encode the cross-architecture assembly instruction sequences by two individual LSTM networks and generate the similarity score based on the distance metric.
	
	\item Redmon \emph{et al.} \cite{redmond2018cross} proposed an instruction pre-training method based on joint objectives. They establish position-based alignments of cross-architecture instruction pairs, and set the mono-architecture and cross-architecture objectives to training instruction embeddings.
	
	%Joint pre-training \cite{redmond2018cross}: Redmon \emph{et al.} proposed an instruction pre-training method based on joint objectives. They establish position-based associations on cross-architecture instruction pairs, and set the mono-architecture and cross-architecture objectives to training instruction representations.
\end{itemize}

\begin{table}
	\caption{Basic block-level cross-architecture binary similarity comparison results on Dataset1.}
	\label{bb_cmp}
	\centering
	\renewcommand{\arraystretch}{1.1}
	%\begin{tabular}{m{110pt}<{\centering}m{40pt}<{\centering}m{45pt}<{\centering}}
	\begin{tabular}{cc}
		\toprule
		%\hline
		Approaches & ROC-AUC \\
		\midrule
		%\hline
		String edit distance & 0.8087 \\
		Char n-gram & 0.7746 \\
		\texttt{Gemini} features + SVM  \cite{xu2017neural} & 0.8647 \\
		%SVM \cite{xu2017neural} & 0.8647 \\
		\texttt{INNEREYE}  \cite{zuo2018neural} & 0.9764 \\
		%LSTM \cite{zuo2018neural} & 0.9764 \\
		Redmond \emph{et al.} \cite{redmond2018cross} & 0.9069 \\
		%Joint pre-training \cite{redmond2018cross} & 0.9069 \\
		
		\midrule
		%\hline
		(- Mono-arch edges) & 0.9917 \\
		(- Cross-archs edges) & 0.9885 \\
		(- Type-specific aggregation) & 0.9825 \\
		
		\midrule
		%\hline
		\textbf{Our approach} & \textbf{0.9924} \\
		%$\triangle$ to the best results of comparison methods & + 1.64\% \\
		\bottomrule
		%\hline
	\end{tabular}
	%\vspace*{-1.2\baselineskip}
\end{table}

We use the instruction embedding files of \texttt{INNEREYE} \cite{innereye_data} and Redmond et al. \cite{cross_data} to make fair comparison. The AUC-ROC on Dataset1 are shown in the upper part of Table \ref{bb_cmp}. From the results, we can mainly draw the following conclusions:

\begin{itemize}
	\item The performance of edit distance, char n-gram, and  \texttt{Gemini} + SVM is not ideal, which shows the syntax or statistical-features based comparison can not handle the significant differences of cross-architecture instruction sets.
	
	\item \texttt{INNEREYE} and our approach outperform other methods by large margins, which proves that exploiting deep neural networks to automatically extract the features of disassembly instruction sequences is a very effective solution.
	
	\item Our approach can achieve the best performance. The improvement is mainly because our graph-based instruction association module can effectively bridge the gap in the representation spaces of cross-architecture instructions. Furthermore, we do not rely on external pre-training to generate initial instruction embeddings, avoiding the additional workload and time consumption.
\end{itemize}

\subsubsection{Ablation studies.}
We design three ablation variants of our approach to evaluate the
core components of our graph-based instruction association module:
(\romannumeral1) Remove the mono-architecture association edges ($e_0$ and $e_1$). 
(\romannumeral2) Remove the cross-architecture association edges ($e_2 \sim e_5$). 
(\romannumeral3) Remove the type-specific neighbor message aggregation mechanism of R-GCN layers and use the classic graph convolutional network \cite{kipf2016semi}. The lower part of Table \ref{bb_cmp} presents the results of the ablation experiments. It proves that the mono-architecture and cross-architecture associations all contribute to performance improvement. And separately aggregating neighbor messages from different types of dependencies can more effectively refine the instruction representation and improve the cross-architecture similarity comparison performance.

%%%%%%%%%%%%%%%%%%%%%%%%%%%%%%%%%%%%%%%%%%%%%%%%%%%%%%%%%%%%%%%%%%%%%%%%%%

\subsection{Function-level Experiments}
In this section, we evaluate our approach at the function-level Dataset2 and compare it with two existing approachs. We set up the binary function search task on Dataset2. The first work we make comparison with is  \texttt{INNEREYE}, which performs best in the basic block-level experiments. The second is another function-level binary similarity comparison approach, \texttt{SAFE} \cite{massarelli2019safe}. It also used \emph{word2vec} to implement instruction pre-training, and deployed the Bi-GRU network to encode disassembly instruction sequences. Then it added the self-attention mechanism to generate instruction representation weights. Instructions that are more important for binary semantic comparison results will be assigned higher weights.

Since the pre-trained  instruction vocabulary provided by \texttt{INNEREYE} are pre-trained on the code corpus compiled by \texttt{clang}, while Dataset2 also involves the \texttt{GCC} compiler, 42.12\% of the instructions are 
out-of-vocabulary (\emph{OOV}). Meanwhile, the instruction embeddings provided by \texttt{SAFE} are also generated by the external corpus inconsistent with the compilation environment of Dataset2. Therefore, we use the trainable instruction embedding and concatenate the char-level features similar to our approach for their instruction vectorization. This phenomenon also illustrates the limitations of the external instruction pre-training approach. Since it is difficult to accurately predict the binary compilation settings in practical applications, serious instruction \emph{OOV} issue is prone to occur.

\begin{table}
	\caption{Function-level cross-architecture binary similarity comparison results on Dataset2.}
	\label{dataset2res}
	\centering
	\renewcommand{\arraystretch}{1.1}
	%begin{tabular}{m{110pt}<{\centering}m{40pt}<{\centering}m{45pt}<{\centering}}
	\begin{tabular}{ccc}
		\toprule
		%\hline
		Approach & Precision@1 & MRR \\
		
		\midrule
		%\hline
		 \texttt{INNEREYE} (42.12\% \emph{OOV}) & 0.5014 & 0.6619 \\
		%Char-features improved-\texttt{INNEREYE} & 0.6386 & 0.7665 \\
		 \texttt{INNEREYE} + Char-features & 0.6386 & 0.7665 \\
		%Char-features improved-\texttt{SAFE} & 0.7097 & 0.7828 \\
		\texttt{SAFE} + Char-features & 0.7097 & 0.7828 \\
		
		%\midrule
		%\hline
		%(- Mono-arch edges) & 0.9440 & 0.9660 \\
		%(- Cross-archs edges) & 0.8992 & 0.9431 \\
		%(- Relational weights) & 0.9244 & 0.9538 \\
		
		\midrule
		%\hline
		\textbf{Our approach} & \textbf{0.9216} & \textbf{0.9550} \\
		%$\triangle$ to the best results of comparison methods & + 29.86\% & + 22.00\% \\
		\bottomrule
		\hline
	\end{tabular}
	%\vspace*{-1.0\baselineskip}
\end{table}

Table \ref{dataset2res} shows the comparison results on Dataset2. From the table, we can see that our approach has significant performance advantages, with at least 21.19 points improvement of the precision@1 value and 17.22 points improvement of the MRR value than the variants of \texttt{INNEREYE} and \texttt{SAFE}. Compared with the results of basic block-level experiments, our function-level improvement is more prominent. On the one hand, the reason is that the instruction sequences of function-level binary snippets are longer and implement more complex semantic functional modules, which increases the difficulty of cross-architecture binary semantic comparison. As shown in Table \ref{dataset_info}, our instruction association graph will establish more multi-type edges on Dataset2, and the effect of bridging the semantic representation space of binary snippets of different architectures is more significant. On the other hand, the function search task needs to compare the query function with multiple candidates and rank the similarity scores, which is more challenging than the pairwise comparison conducted at basic block-level.

\subsection{Real-world IoT Malware Reuse Function Matching Experiments}
In this section, we use Dataset3 to evaluate the scalability and practicability of our method in real-world applications. Dataset3 is constructed from malware captured in the public network by IoT honeypots. We use \emph{radare2} to process unstripped malware samples of \emph{ARM} and \emph{MIPS} architectures and construct a large-scale dataset containing over 460k cross-architecture function pairs, characterizing malicious behaviors from a more refined perspective. We also make comparisons with \texttt{INNEREYE} and \texttt{SAFE}. Since their pre-trained instruction embeddings did not support \emph{MIPS} architecture, we still compare with their variants using trainable embeddings and char-level features for instruction vectorization.

\begin{table*}
	\caption{Cross-architecture binary similarity comparison results on Dataset3.}
	%\caption{Cross-architecture reuse function matching results on Dataset3}
	\label{dataset3res}
	\centering
	\renewcommand{\arraystretch}{1.1}
	\begin{tabular}{m{108pt}<{\centering}m{45pt}<{\centering}m{45pt}<{\centering}m{70pt}<{\centering}m{78pt}<{\centering}}
		\toprule
		%\hline
		Approach & Precision@1 & MRR & Offline training-time (seconds/epoch) & Online prediction-time (milliseconds/pair) \\
		
		\midrule
		%\hline
		%Improved-\texttt{INNEREYE} & 0.7780 & 0.8366 & 301 & 0.58 \\
		
		\texttt{INNEREYE} + Char-features  & 0.7780 & 0.8366 & 101 & 1.58 \\
		%205	0.703
		%\texttt{INNEREYE} + Char-features  & 0.7780 & 0.8366 & 2.81 & 1.58 \\
		%Improved-\texttt{SAFE} & 0.7840 & 0.8386 & 318 & 0.84  \\
		\texttt{SAFE}  + Char-features& 0.7840 & 0.8386 & 318 & 1.84  \\
		%220	0.20
		%\texttt{SAFE}  + Char-features& 0.7840 & 0.8386 & 8.83 & 1.84  \\
		%\hline
		\textbf{Our approach} & \textbf{0.9375} & \textbf{0.9597} & 823 & 3.72 \\
		%647	5.63
		%textbf{Our approach} & \textbf{0.9375} & \textbf{0.9597} & 22.86 & 3.72 \\
		\bottomrule
		%\hline
	\end{tabular}
	%\vspace*{-1.0\baselineskip}
\end{table*}

Table \ref{dataset3res} shows the performance of the three approaches on Dataset3, along with their offline training time (seconds per epoch) and online prediction time (milliseconds per function pair). The results show that our method significantly outperforms the variants of \texttt{INNEREYE} and \texttt{SAFE}, with precision@1 and MRR values improving at least 16.97 and 12.11 points, respectively. Due to the instruction association graph construction and the R-GCN-based multi-type dependencies aggregation on the graph, the training time and prediction time of our method are increased. However, the online prediction speed is still acceptable as inferencing the similarity score of a binary function pair can be completed in 3.72 milliseconds. Meanwhile, we do not need additional time collecting external code corpus consistent with the specific application's compilation settings and pre-training disassembly instruction embeddings. In conclusion, our method can be effectively applied to large-scale real-world cross-architecture binary similarity comparison collection and provides a valuable solution for defending against malware reused and propagated on IoT devices of different architectures.

\section{Related Work}
\subsection{Traditional Binary Similarity Comparison Approaches}
Traditional binary similarity comparison approaches are mainly divided into two categories, static analysis techniques comparing syntax or structural features \cite{lee2013function,khoo2013rendezvous,cesare2013control,luo2014semantics,pewny2014leveraging,ruttenberg2014identifying,qiao2016fast,huang2017binsequence} and dynamic analysis techniques comparing behavioral semantics \cite{lindorfer2012lines,ming2015memoized,ming2017binsim,kargen2017towards,wang2017memory}. For static analysis-based methods, \texttt{Rendezvous} \cite{khoo2013rendezvous} extracted the instruction n-perms, control flow sub-graphs, and constants of binary code to construct a code search engine. Qiao \emph{et al.} \cite{qiao2016fast} used the \emph{simhash} algorithm to generate basic block signatures, then exploited inverted index to achieve fast reuse function detection. \texttt{CoP} \cite{luo2014semantics} first checked block-level semantic equivalence with theorem prover, and then performed the breadth-first search on the inter-procedural control-flow-graph (I-CFG) to compute path-level binary semantic similarity. These approaches are designed for mono-architecture binary comparison without handling the differences in instruction sets of different architectures. For dynamic analysis-based methods, Ulf \emph{et al.} \cite{kargen2017towards} recorded the execution traces and output values of a binary pair when given same input, and used matching features to identify programs with similar semantics but different syntax characteristics. \texttt{BinSim} \cite{ming2017binsim} performed enhanced dynamic slicing and extracted the symbolic formulas of the code fragments to check the semantic equivalence. Dynamic analysis will meet challenges when analyzing large binary collections, and it requires additional efforts to configure execution environments supporting diverse architectures and compilation settings.

To support cross-architecture comparison, some existing work converted the instructions of different architectures into the intermediate representation (IR) \cite{pewny2015cross,hu2016cross,chandramohan2016bingo,feng2017extracting}. Specifically, \texttt{Multi}-\texttt{MH} \cite{pewny2015cross} first converted binary code into platform-independent VEX-IR, then used the input-output pairs of basic blocks to construct their semantic signatures. \texttt{XMATCH} \cite{feng2017extracting} conducted static analysis on IR and extracted conditional formulas as semantic features, improving the binary vulnerability search accuracy. These methods may be limited by the adaptability and precision of IR extraction tools. \texttt{Esh} \cite{david2016statistical} decomposed binary procedures into small fragments, named strands, then checked the semantically equivalence of strand pairs and used statistical reasoning for procedure-level similarity comparison, but it also suffers the unscalable issue on the large-scale collections.

\subsection{Learning-based Binary Similarity Comparison Approaches}
Recently, deep learning technology has achieved promising improvements in intelligent program analysis. To improve the effectiveness and efficiency of the binary similarity comparison task, researchers pay attention to learning-based approaches \cite{eschweiler2016discovre,feng2016scalable,lageman2016b,xu2017neural,zuo2018neural,liu2018alphadiff,gao2018vulseeker,massarelli2019safe,ding2019asm2vec,massarelli2019investigating,yu2020order,liang2020fit,duan2020deepbindiff}. \texttt{$\alpha$-diff} exploited the siamese-CNN network to achieve cross-version binary code similarity detection. They adopted in-batch random negative sampling and contrastive loss to optimize the model. \texttt{Asm2Vec} \cite{ding2019asm2vec} used the PV-DM model to implement binary function vectorization, achieving accurate binary clone detection against changes introduced by code obfuscation and optimization techniques. 

For cross-architecture binary similarity comparison, \texttt{Gemini} \cite{xu2017neural} extracted the attribute control flow graph (ACFG) of the binary functions. Then it used the \emph{structure2vec} network and cosine similarity to implement bug search for IoT firmware images. \texttt{VulSeeker} \cite{gao2018vulseeker} extends ACFG into labelled semantic flow graph (LSFG) by adding additional data flow edges. \texttt{INNEREYE} \cite{zuo2018neural} leveraged \emph{wordvec} to pre-train instruction embeddings on external code corpus, then implemented binary comparison by LSTM encoders and Manhattan distance. \texttt{SAFE} \cite{massarelli2019safe} added the self-attention mechanism to calculate the importance weights of the instructions within the disassembled code sequence, improving the cross-architecture binary similarity matching accuracy. The common dilemma of existing learning-based approaches is that the cross-architecture binary snippets are encoded separately by neural layers and then compared based on the similarity metrics. The significant difference between cross-architecture instruction sets will lead to a considerable gap in their semantical representation spaces, which limits the comparison performance. Our method utilizes the multi-relational instruction association graph and R-GCN layers to close the gap in the semantic spaces of cross-architecture instructions, effectively alleviating the deficiency.

\section{Conclusion}
In this paper, we propose a novel cross-architecture binary similarity comparison approach. We design an instruction association graph schema to bridge the gap in the semantic spaces of instruction sets from different architectures. It consists of six types of dependencies, which respectively define the context relevance of mono-architecture instruction tokens and multi-perspective semantic associations of cross-architecture tokens. We leverage the R-GCN network to propagate the multi-type dependencies within the graph and improve the cross-architecture binary matching performance.
We conduct extensive experiments on datasets of different granularities. Results show that it outperforms existing learning-based approaches on basic block-level and function-level comparisons. Furthermore, our approach can achieve effective cross-architecture reuse function detection on a large-scale IoT malware dataset collected from the real-world network environment, which is meaningful for identifying malware spread on IoT devices of various architectures and defending against IoT attacks.

%\scriptsize
\bibliographystyle{splncs04}
\bibliography{reference}

\begin{thebibliography}{10}
\providecommand{\url}[1]{\texttt{#1}}
\providecommand{\urlprefix}{URL }
\providecommand{\doi}[1]{https://doi.org/#1}

\bibitem{antonakakis2017understanding}
Antonakakis, M., April, T., Bailey, M., Bernhard, M., Bursztein, E., Cochran,
  J., Durumeric, Z., Halderman, J.A., Invernizzi, L., Kallitsis, M., et~al.:
  Understanding the mirai botnet. In: 26th USENIX security symposium (USENIX
  Security 17). pp. 1093--1110 (2017)

\bibitem{cesare2013control}
Cesare, S., Xiang, Y., Zhou, W.: Control flow-based malware variantdetection.
  IEEE Transactions on Dependable and Secure Computing  \textbf{11}(4),
  307--317 (2013)

\bibitem{chandramohan2016bingo}
Chandramohan, M., Xue, Y., Xu, Z., Liu, Y., Cho, C.Y., Tan, H.B.K.: Bingo:
  Cross-architecture cross-os binary search. In: Proceedings of the 2016 24th
  ACM SIGSOFT International Symposium on Foundations of Software Engineering.
  pp. 678--689 (2016)

\bibitem{conneau2017supervised}
Conneau, A., Kiela, D., Schwenk, H., Barrault, L., Bordes, A.: Supervised
  learning of universal sentence representations from natural language
  inference data. In: Proceedings of the 2017 Conference on Empirical Methods
  in Natural Language Processing, {EMNLP} 2017, Copenhagen, Denmark, September
  9-11, 2017 (2017)

\bibitem{costin2018iot}
Costin, A., Zaddach, J.: Iot malware: Comprehensive survey, analysis framework
  and case studies. BlackHat USA  \textbf{1}(1), ~1--9 (2018)

\bibitem{cozzi2020tangled}
Cozzi, E., Vervier, P.A., Dell'Amico, M., Shen, Y., Bilge, L., Balzarotti, D.:
  The tangled genealogy of iot malware. In: Annual Computer Security
  Applications Conference. pp. 1--16 (2020)

\bibitem{david2016statistical}
David, Y., Partush, N., Yahav, E.: Statistical similarity of binaries. ACM
  SIGPLAN Notices  \textbf{51}(6),  266--280 (2016)

\bibitem{ding2019asm2vec}
Ding, S.H., Fung, B.C., Charland, P.: Asm2vec: Boosting static representation
  robustness for binary clone search against code obfuscation and compiler
  optimization. In: 2019 IEEE Symposium on Security and Privacy (SP). pp.
  472--489. IEEE (2019)

\bibitem{duan2020deepbindiff}
Duan, Y., Li, X., Wang, J., Yin, H.: Deepbindiff: Learning program-wide code
  representations for binary diffing. In: Proceedings of the 27th Annual
  Network and Distributed System Security Symposium (NDSS’20) (2020)

\bibitem{eschweiler2016discovre}
Eschweiler, S., Yakdan, K., Gerhards-Padilla, E.: discovre: Efficient
  cross-architecture identification of bugs in binary code. In: NDSS (2016)

\bibitem{farhadi2014binclone}
Farhadi, M.R., Fung, B.C., Charland, P., Debbabi, M.: Binclone: Detecting code
  clones in malware. In: 2014 Eighth International Conference on Software
  Security and Reliability (SERE). pp. 78--87. IEEE (2014)

\bibitem{feng2017extracting}
Feng, Q., Wang, M., Zhang, M., Zhou, R., Henderson, A., Yin, H.: Extracting
  conditional formulas for cross-platform bug search. In: Proceedings of the
  2017 ACM on Asia Conference on Computer and Communications Security. pp.
  346--359 (2017)

\bibitem{feng2016scalable}
Feng, Q., Zhou, R., Xu, C., Cheng, Y., Testa, B., Yin, H.: Scalable graph-based
  bug search for firmware images. In: Proceedings of the 2016 ACM SIGSAC
  Conference on Computer and Communications Security. pp. 480--491 (2016)

\bibitem{FeyLenssen2019}
Fey, M., Lenssen, J.E.: Fast graph representation learning with {PyTorch
  Geometric}. In: ICLR Workshop on Representation Learning on Graphs and
  Manifolds (2019)

\bibitem{gao2018vulseeker}
Gao, J., Yang, X., Fu, Y., Jiang, Y., Sun, J.: Vulseeker: a semantic learning
  based vulnerability seeker for cross-platform binary. In: 2018 33rd IEEE/ACM
  International Conference on Automated Software Engineering (ASE). pp.
  896--899. IEEE (2018)

\bibitem{herwig2019measurement}
Herwig, S., Harvey, K., Hughey, G., Roberts, R., Levin, D.: Measurement and
  analysis of hajime, a peer-to-peer iot botnet. In: Network and Distributed
  Systems Security (NDSS) Symposium (2019)

\bibitem{hu2013mutantx}
Hu, X., Shin, K.G., Bhatkar, S., Griffin, K.: Mutantx-s: Scalable malware
  clustering based on static features. In: 2013 $\{$USENIX$\}$ Annual Technical
  Conference ($\{$USENIX$\}$$\{$ATC$\}$ 13). pp. 187--198 (2013)

\bibitem{hu2016cross}
Hu, Y., Zhang, Y., Li, J., Gu, D.: Cross-architecture binary semantics
  understanding via similar code comparison. In: 2016 IEEE 23rd International
  Conference on Software Analysis, Evolution, and Reengineering (SANER).
  vol.~1, pp. 57--67. IEEE (2016)

\bibitem{huang2017binsequence}
Huang, H., Youssef, A.M., Debbabi, M.: Binsequence: Fast, accurate and scalable
  binary code reuse detection. In: Proceedings of the 2017 ACM on Asia
  Conference on Computer and Communications Security. pp. 155--166 (2017)

\bibitem{kargen2017towards}
Karg{\'e}n, U., Shahmehri, N.: Towards robust instruction-level trace alignment
  of binary code. In: 2017 32nd IEEE/ACM International Conference on Automated
  Software Engineering (ASE). pp. 342--352. IEEE (2017)

\bibitem{khoo2013rendezvous}
Khoo, W.M., Mycroft, A., Anderson, R.: Rendezvous: A search engine for binary
  code. In: 2013 10th Working Conference on Mining Software Repositories (MSR).
  pp. 329--338. IEEE (2013)

\bibitem{kipf2016semi}
Kipf, T.N., Welling, M.: Semi-supervised classification with graph
  convolutional networks  (2016)

\bibitem{lageman2016b}
Lageman, N., Kilmer, E.D., Walls, R.J., McDaniel, P.D.: B in dnn: Resilient
  function matching using deep learning. In: International Conference on
  Security and Privacy in Communication Systems. pp. 517--537. Springer (2016)

\bibitem{lee2013function}
Lee, Y.R., Kang, B., Im, E.G.: Function matching-based binary-level software
  similarity calculation. In: Research in Adaptive and Convergent Systems,
  RACS'13, Montreal, QC, Canada, October 1-4, 2013. pp. 322--327. {ACM} (2013)

\bibitem{edit_distance}
python Levenshtein: https://pypi.org/project/python-Levenshtein/

\bibitem{liang2020fit}
Liang, H., Xie, Z., Chen, Y., Ning, H., Wang, J.: Fit: Inspect vulnerabilities
  in cross-architecture firmware by deep learning and bipartite matching.
  Computers \& Security  \textbf{99},  102032 (2020)

\bibitem{lindorfer2012lines}
Lindorfer, M., Di~Federico, A., Maggi, F., Comparetti, P.M., Zanero, S.: Lines
  of malicious code: Insights into the malicious software industry. In:
  Proceedings of the 28th Annual Computer Security Applications Conference. pp.
  349--358 (2012)

\bibitem{liu2018alphadiff}
Liu, B., Huo, W., Zhang, C., Li, W., Li, F., Piao, A., Zou, W.: $\alpha$diff:
  cross-version binary code similarity detection with dnn. In: Proceedings of
  the 33rd ACM/IEEE International Conference on Automated Software Engineering.
  pp. 667--678 (2018)

\bibitem{luo2014semantics}
Luo, L., Ming, J., Wu, D., Liu, P., Zhu, S.: Semantics-based
  obfuscation-resilient binary code similarity comparison with applications to
  software plagiarism detection. In: Proceedings of the 22nd ACM SIGSOFT
  International Symposium on Foundations of Software Engineering. pp. 389--400
  (2014)

\bibitem{massarelli2019investigating}
Massarelli, L., Di~Luna, G.A., Petroni, F., Querzoni, L., Baldoni, R.:
  Investigating graph embedding neural networks with unsupervised features
  extraction for binary analysis. In: Proceedings of the 2nd Workshop on Binary
  Analysis Research (BAR) (2019)

\bibitem{massarelli2019safe}
Massarelli, L., Di~Luna, G.A., Petroni, F., Baldoni, R., Querzoni, L.: Safe:
  Self-attentive function embeddings for binary similarity. In: International
  Conference on Detection of Intrusions and Malware, and Vulnerability
  Assessment. pp. 309--329. Springer (2019)

\bibitem{ming2017binsim}
Ming, J., Xu, D., Jiang, Y., Wu, D.: Binsim: Trace-based semantic binary
  diffing via system call sliced segment equivalence checking. In: 26th
  $\{$USENIX$\}$ Security Symposium ($\{$USENIX$\}$ Security 17). pp. 253--270
  (2017)

\bibitem{ming2015memoized}
Ming, J., Xu, D., Wu, D.: Memoized semantics-based binary diffing with
  application to malware lineage inference. In: IFIP International Information
  Security and Privacy Conference. pp. 416--430. Springer (2015)

\bibitem{ng2013expose}
Ng, B.H., Prakash, A.: Expose: Discovering potential binary code re-use. In:
  2013 IEEE 37th Annual Computer Software and Applications Conference. pp.
  492--501. IEEE (2013)

\bibitem{pewny2015cross}
Pewny, J., Garmany, B., Gawlik, R., Rossow, C., Holz, T.: Cross-architecture
  bug search in binary executables. In: 2015 IEEE Symposium on Security and
  Privacy. pp. 709--724. IEEE (2015)

\bibitem{pewny2014leveraging}
Pewny, J., Schuster, F., Bernhard, L., Holz, T., Rossow, C.: Leveraging
  semantic signatures for bug search in binary programs. In: Proceedings of the
  30th Annual Computer Security Applications Conference. pp. 406--415 (2014)

\bibitem{qiao2016fast}
Qiao, Y., Yun, X., Zhang, Y.: Fast reused function retrieval method based on
  simhash and inverted index. In: 2016 IEEE Trustcom/BigDataSE/ISPA. pp.
  937--944. IEEE (2016)

\bibitem{radare2}
radare2: https://www.radare.org/n/radare2.html

\bibitem{cross_data}
Redmond, K., Luo, L., Zeng, Q.:
  https://github.com/nlp-code-analysis/cross-arch-instr-model/

\bibitem{redmond2018cross}
Redmond, K., Luo, L., Zeng, Q.: A cross-architecture instruction embedding
  model for natural language processing-inspired binary code analysis. arXiv
  preprint arXiv:1812.09652  (2018)

\bibitem{ruttenberg2014identifying}
Ruttenberg, B., Miles, C., Kellogg, L., Notani, V., Howard, M., LeDoux, C.,
  Lakhotia, A., Pfeffer, A.: Identifying shared software components to support
  malware forensics. In: International Conference on Detection of Intrusions
  and Malware, and Vulnerability Assessment. pp. 21--40. Springer (2014)

\bibitem{scarselli2008graph}
Scarselli, F., Gori, M., Tsoi, A.C., Hagenbuchner, M., Monfardini, G.: The
  graph neural network model. IEEE transactions on neural networks
  \textbf{20}(1),  61--80 (2008)

\bibitem{schlichtkrull2018modeling}
Schlichtkrull, M., Kipf, T.N., Bloem, P., Berg, R.v.d., Titov, I., Welling, M.:
  Modeling relational data with graph convolutional networks. In: European
  semantic web conference. pp. 593--607. Springer (2018)

\bibitem{wang2020iotcmal}
Wang, B., Dou, Y., Sang, Y., Zhang, Y., Huang, J.: Iotcmal: Towards a hybrid
  iot honeypot for capturing and analyzing malware. In: ICC 2020-2020 IEEE
  International Conference on Communications (ICC). pp.~1--7. IEEE (2020)

\bibitem{wang2017memory}
Wang, S., Wu, D.: In-memory fuzzing for binary code similarity analysis. In:
  2017 32nd IEEE/ACM International Conference on Automated Software Engineering
  (ASE). pp. 319--330. IEEE (2017)

\bibitem{xu2017neural}
Xu, X., Liu, C., Feng, Q., Yin, H., Song, L., Song, D.: Neural network-based
  graph embedding for cross-platform binary code similarity detection. In:
  Proceedings of the 2017 ACM SIGSAC Conference on Computer and Communications
  Security. pp. 363--376 (2017)

\bibitem{xu2017spain}
Xu, Z., Chen, B., Chandramohan, M., Liu, Y., Song, F.: Spain: security patch
  analysis for binaries towards understanding the pain and pills. In: 2017
  IEEE/ACM 39th International Conference on Software Engineering (ICSE). pp.
  462--472. IEEE (2017)

\bibitem{yu2020order}
Yu, Z., Cao, R., Tang, Q., Nie, S., Huang, J., Wu, S.: Order matters:
  Semantic-aware neural networks for binary code similarity detection. In:
  Proceedings of the AAAI Conference on Artificial Intelligence. vol.~34, pp.
  1145--1152 (2020)

\bibitem{innereye_data}
Zuo, F., Li, X., Young, P., Luo, L., Zeng, Q., Zhang, Z.:
  https://nmt4binaries.github.io/

\bibitem{zuo2018neural}
Zuo, F., Li, X., Young, P., Luo, L., Zeng, Q., Zhang, Z.: Neural machine
  translation inspired binary code similarity comparison beyond function pairs.
  In: 26th Annual Network and Distributed System Security Symposium, {NDSS}
  2019, San Diego, California, USA, February 24-27, 2019 (2018)

\end{thebibliography}

%\bibliography{iag_reference}
%\bibliographystyle{splncs04}
\end{document}